\documentclass[amsmath,amssymb,aps,pra,twocolumn,showpacs,superscriptaddress]{revtex4-1}

\usepackage {graphicx}
\usepackage{epsf}
\usepackage[caption=false]{subfig}
\usepackage {multirow}
\usepackage{ragged2e}
\usepackage{eqnarray,subeqnarray}

\begin{document}

\title{A Coulomb Over-the-Barrier Monte Carlo Simulation to probe Ion-Dimer Collision Dynamics}

\author{W. Iskandar}
\email{wiskandar@lbl.gov}
\affiliation{Chemical Sciences Division, Lawrence Berkeley National Laboratory, Berkeley, CA-94720, USA}
\author{X. Fl\'echard}
\email{flechard@lpccaen.in2p3.fr}
\affiliation{Normandie Univ, ENSICAEN, UNICAEN, CNRS/IN2P3, LPC Caen, 14000 Caen, France}
\author{J. Matsumoto}
\affiliation{Department of Chemistry, Tokyo Metropolitan University, 1-1 Minamiosawa, Hachiouji-shi, Tokyo 192-0397, Japan}
\author{A. Leredde}
\affiliation{Normandie Univ, ENSICAEN, UNICAEN, CNRS/IN2P3, LPC Caen, 14000 Caen, France}
\affiliation{LPSC, Universit\'e Grenoble Alpes, CNRS/IN2P3, 38026 Grenoble, France}
\author{S. Guillous}
\affiliation{CIMAP, CEA-CNRS-ENSICAEN-Universit\'e de Caen, BP 5133, F-14070 Caen cedex 5, France}
\author{D. Hennecart}
\affiliation{CIMAP, CEA-CNRS-ENSICAEN-Universit\'e de Caen, BP 5133, F-14070 Caen cedex 5, France}
\author{J. Rangama}
\affiliation{CIMAP, CEA-CNRS-ENSICAEN-Universit\'e de Caen, BP 5133, F-14070 Caen cedex 5, France}
\author{A. M\'ery}
\affiliation{CIMAP, CEA-CNRS-ENSICAEN-Universit\'e de Caen, BP 5133, F-14070 Caen cedex 5, France}
\author{B. Gervais}
\affiliation{CIMAP, CEA-CNRS-ENSICAEN-Universit\'e de Caen, BP 5133, F-14070 Caen cedex 5, France}
\author{H. Shiromaru}
\affiliation{Department of Chemistry, Tokyo Metropolitan University, 1-1 Minamiosawa, Hachiouji-shi, Tokyo 192-0397, Japan}
\author{A. Cassimi}
\affiliation{CIMAP, CEA-CNRS-ENSICAEN-Universit\'e de Caen, BP 5133, F-14070 Caen cedex 5, France}

\date{\today}

\begin{abstract}
We present a combined theoretical and experimental study of primary and post-collision mechanisms involved when colliding low energy multiply charged ions with van der Waals dimers.
The collision dynamics is investigated using a classical calculation based on the Coulombic Over-the-Barrier Model adapted to rare-gas dimer targets. Despite its simplicity, the model predictions are found in very good agreement with experimental results obtained using COLd Target Recoil Ion Momentum Spectroscopy (COLTRIMS), both for the relative yields of the different relaxation processes and for the associated transverse momentum exchange distributions between the projectile and the target. This agreement shows to which extent van der Waals dimers can be assimilated to independent atoms.
\end{abstract}

\pacs{34.70.+e, 36.40.Mr, 34.10.+x, 32.80.Hd, 36.40.-c}

\maketitle
\section{\label{sec1}INTRODUCTION}
The complete understanding of primary and post-collision mechanisms involving ions, atoms and molecules is of primordial importance in many domains of science. These processes generate interest in fundamental physics as well as in interdisciplinary research such as astrophysics \cite{Dennerl10}, accelerator technologies, fusion plasma physics \cite{Logan08}, biological and medical treatments \cite{Amaldi05,Nikoghosyan04}. For decades, many theoretical and experimental investigations have thus been devoted to the understanding of collision mechanisms particularly for ionic projectiles colliding with atomic targets.
The relative cross sections associated to the different elementary processes (ionization, excitation and electron capture from the atomic target) depend strongly on the collision regime, given by the comparison of the projectile velocity ($v_p$) to the orbital velocity of active electrons ($v_e$) and by the collision asymmetry, i.e. the ratio between the projectile and target atomic numbers \cite{DuBois86,Shingal87,Knoop05,Flechard15}. It is today well established that, as ionization and excitation of the target dominates in the high velocity regime, charge exchange (or electron capture) is by far the most probable process at low energy. For the latter, concomitant advances in experimental and theoretical techniques have led to a quite complete knowledge of the collision dynamics, which is reviewed in textbooks such as \cite{Bransden92}.

Low energy collisions of ionic projectiles with molecular targets such as diatomic molecules are more complex and their investigation is still a theoretical and experimental challenge. The complexity is due to additional degrees of freedom such as the orientation of the molecule and to the multiple interactions between all the components of the molecule (electrons and nuclei). It opens up new primary mechanisms, such as the screening effect of both centers active electrons during the collision, and new post-collision mechanisms such as energy and/or electron exchange between the two sites of the molecule, leading to new relaxation processes.

Primary and post-collision mechanisms can also strongly depend on the bond type of the molecular edifice such as in covalently bound molecules or van der Waals (vdW) bound dimers. When an ionic projectile ionizes or captures many electrons from a covalent molecule, the latter tends to dissociate into equally charged fragments due to the delocalization of the valence electrons throughout the molecule \cite{Itzhak93,Ehrich02}. In contrast, for vdW bound dimers, the long separation distance between the two atoms of the dimer may lead to weaker charge rearrangement. Non-equally charged fragmentation becomes thus more efficient and can even dominate the equally charged channel \cite{Matsumoto10}.
Furthermore, in the low energy collision regime, the long separation distance and the weak charge mobility between the two sites of dimers lead to projectiles preferentially scattered in the direction of the most charged fragment of the dimer \cite{Iskandar14}, as opposed to what was previously observed with N$_2$ molecular targets \cite{Ehrich02}. Another specificity of vdW bound dimers when compared to covalent molecules is the appearance of new post-collision mechanisms. Many results have shown that producing non-equally charged atoms of the dimer leads to new relaxation processes such as Radiative Charge Transfer (RCT) and Interatomic/Intermolecular Coulombic Decay (ICD) involving electron and/or energy transfer between one neutral site and one charged site of the dimer \cite{Cederbaum97,Jahnke04,Matsumoto10,Kim13,Iskandar15}.

For a better understanding of these primary and post-collision mechanisms related to vdW dimer targets, several theoretical approaches can be considered. A full quantum-mechanical treatment of collisions between multiply charged ion (MCI) projectiles and molecular targets requires solving the time dependent Schr\"odinger equation for a given state by taking into account the kinetic and potential energies of all partners of the collision. Such an ab initio approach is impossible in practice and approximation methods have to be used. Classical models are known to provide general insight into the interaction mechanisms and dependencies on the collision parameters. For atomic targets, a simple classical model valid for one-electron capture was introduced by Bohr and Lindhard \cite{Bohr54} and later elaborated by Knudsen et al. \cite{Knudsen81}. Here, an electron can only be transferred if the Coulomb force exerted by the projectile exceeds the initial binding force and if the electron kinetic energy in the projectile frame is smaller than its potential energy. The model predicts an energy independent cross section at low impact energies. At high energies, the cross section decreases as $v_p^{-7}$, in good agreement with experimental data \cite{Knudsen81}. Later on, interest in multiple electron processes led to the birth of the Classical Over-the-Barrier Model (COBM). The COBM, mainly developed by Ryufuku et al \cite{Ryufuku80}, Mann et al \cite{Mann81}, Barany et al \cite{Barany85} and Niehaus \cite{Niehaus86}, is based on the idea that electrons can transit from the target to the projectile at given internuclear distances for which the height of the potential barrier between the two nuclei is lower than the Stark shifted binding energy of the electrons. This model has been successfully used for MCI projectiles colliding atomic targets in the low energy regime ($v_p$ ranging from 0.01 to 1 a.u., typically) to predict cross sections associated to the primary collision processes \cite{Niehaus86,Niehaus90} as well as to post-collision processes \cite{Niehaus88,Guillemot90}.

For molecular targets, a three-center COBM based on the model described by Barany et al \cite{Barany85} has recently been developed by Ichimura and Ohyama-Yamaguchi \cite{Yamaguchi09}. This model was used  to investigate multiple electron capture from diatomic molecules in collisions with slow MCIs.
The results were found to be consistent with experimental results for N$_2$ molecules \cite{Ehrich02}, showing a dominant dissociation into equally charged fragments for even capture multiplicity or with only one electron difference between the two charged fragments for odd numbers of electron capture \cite{Yamaguchi09}. Moreover, the calculation reproduced the higher probability to end up with the most charged fragment located on the molecule farther site from the projectile, due to the strong polarization of the N$_2$ molecule in the presence of the MCIs \cite{Yamaguchi05,Yamaguchi09}.
Beside the treatment of covalent molecular targets, the authors have applied the same analytical methodology to rare gas dimer targets by adding an adjustable electron screening parameter \cite{Yamaguchi13}. The calculations have shown an increase of charge asymmetry in the ion pair distribution when increasing the screening parameter.

For further investigation of the collision dynamics with atomic dimer targets, we have used a similar approach but including additional ingredients. To perform a more complete treatment of the collision, our model comprises two distinct stages, the way-in and the way-out, as proposed by Niehaus for atomic targets \cite{Niehaus86}. Secondly, we used Monte Carlo (MC) simulations to facilitate both the theoretical treatment and the comparison with the experimental data. These developments allow the prediction of the final ion pair production, give access to capture multiplicity on each site as a function of the impact parameter $\vec{b}$ in the molecular frame, and provide the transverse momentum exchange between the projectile and each center of the dimer all along the interaction path. In our previous papers \cite{Iskandar15,Iskandar14}, the predictions of this model have already been successfully compared to experimental results, leading to a better understanding of primary and post-collision mechanisms involved with dimer targets. Only few details were provided on the calculations and the model itself. In the following sections, a complete description of the different steps of the MC-COBM calculations is given, followed by a comparison with experimental data obtained with projectile velocities ranging from $v_p \sim 0.3$~to~$ 0.4$ a.u..
\section{\label{sec2}Classical Over-the-Barrier Model}
For the sake of simplification, we will start in Sec.\ref{sec2-1} by presenting the MC-COBM for atomic targets. The method is very similar to the one initially proposed by Niehaus \cite{Niehaus86} as charge exchange probabilities are calculated using the same ingredients. However, it includes new features such as randomly generated trajectories of the projectiles instead of an integrated cross section calculation, and an estimation of the transverse momentum exchange. We also introduce the notion of effective charge of the projectile and of the target to provide a simplified formulation of the equations found in~\cite{Niehaus86}. The adaptation of this model to an atomic dimer target is then presented in Sec.\ref{sec2-2}. 
\subsection{\label{sec2-1}Atomic Target}
\subsubsection{\label{sec2-1-0}Principle of the COBM}
\begin{figure*}
\includegraphics[width=2.\columnwidth,clip]{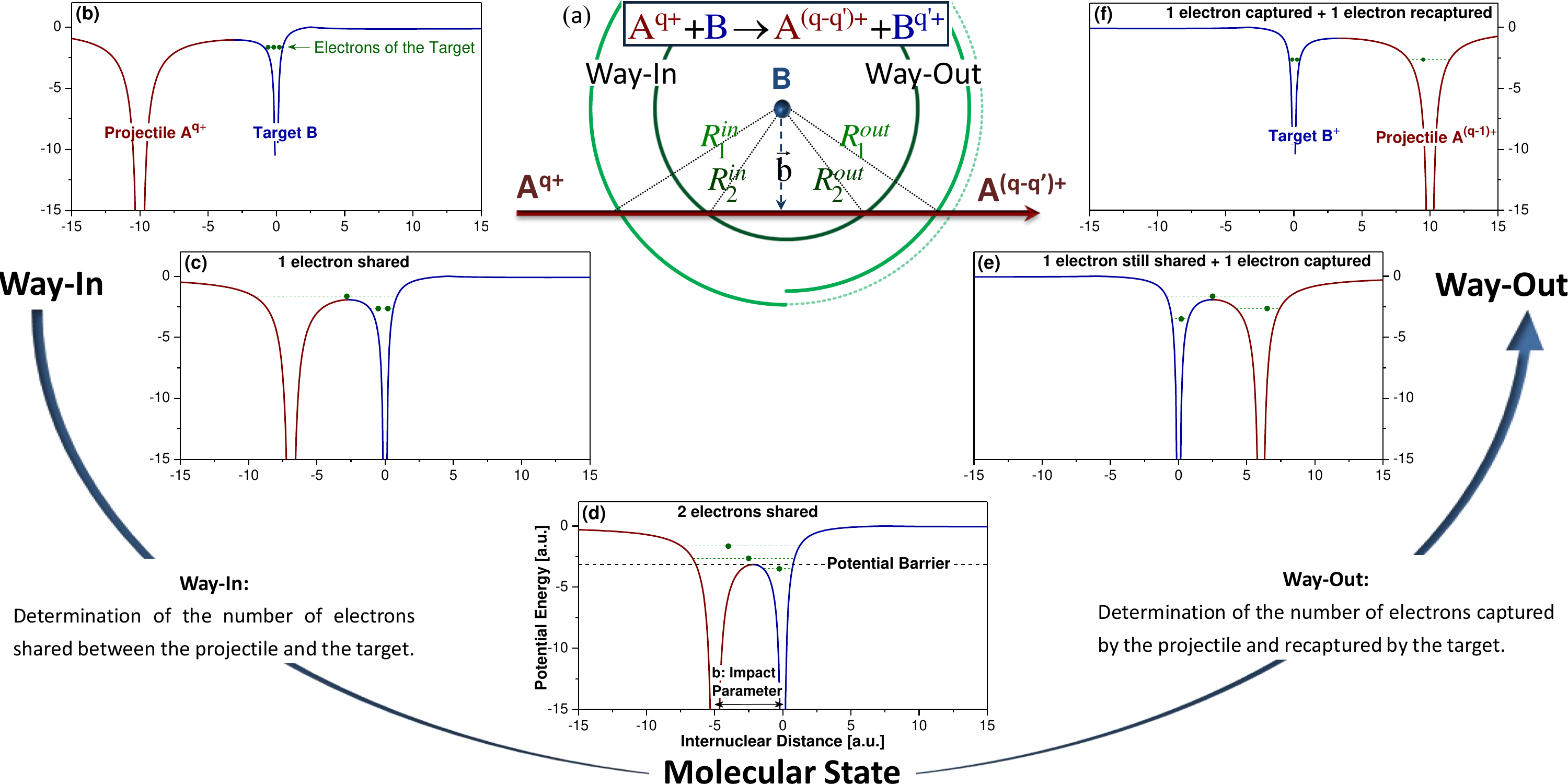}
\caption{(a) Schematic view of the interaction between a projectile A$^{q+}$ and a target B with $N_{B}$ shared electrons involved in the way-in ($N_{B}$=2 in this example), as the projectile is approaching the target, and where 0 to $N_B$ electrons may be captured by the projectile in the way-out, as the projectile is leaving the target. (b to f) Schematic representation of the potential barrier evolution in the way-in and in the way-out of the collision. In the way-in, the approach of the projectile towards the target lowers the potential barrier height between them (b). At a critical internuclear distance called sharing radius $R_i^{in}$ for each electron $i$, the potential barrier may become lower than the electron binding energy on the target (c), causing this electron to occupy a molecular state induced by the proximity of the projectile and the target. The maximum number of electrons shared will be reached at the impact parameter distance $||$$\vec{b}$$||$ where the way-in stage ends and the way-out stage starts (d). In the way-out, the projectile will leave the target causing an increase of the potential barrier height. For each shared electron $i$, at a given distance called capture radius $R_i^{out}$, the model will determine the probability for this active electron to be captured by the projectile or to be recaptured by the target (e, f).}
\label{COBM}
\end{figure*}

Within the COBM model, the collision of an ionic projectile, noted A$^{q+}$, with an atomic target, noted B, is processed in two distinct parts that will be referred to as the way-in and the way-out (Fig.\ref{COBM}). The way-in and the way-out correspond respectively to the stage with the projectile approaching the target and to the stage with the projectile moving away from the target. In the model, the projectile trajectory is approximated by a straight line.

On the way-in, as the projectile gets closer to the target, the Coulomb potential barrier between the target and the projectile decreases with the internuclear separation of the two partners of the collision. An electron is thus transferred to a molecular state when the potential barrier becomes lower than its initial binding energy to the target. This transfer occurs at critical internuclear distances between the projectile and the target called sharing radii. Each electron, numbered $i$ ($i$ increasing with the binding energy of the electron), has its own sharing radius $R_i^{in}$ depending on its initial binding energy to the target $I^{B}_i$ and on the projectile and target effective charges $q^{A_{in}}_i$ and $q^{B_{in}}_i$:
\begin{eqnarray}
	R^{in}_{i} &=& \frac{i+2\sqrt{q^{A_{in}}_i\times q^{B_{in}}_i}}{I^{B}_i}\label{R_in}
\end{eqnarray}

In the COBM model, shared electrons are transferred to molecular states where they do not contribute to the charge screening between the projectile and the target nuclei. This means that in the way-in, the projectile effective charge  keeps its initial value $q^{A_{in}}_i=q$ while the target effective charge $q^{B_{in}}_i=i$ increases by one unit each time one electron is shared. Taking into account the Coulomb interaction with the projectile, the binding energy of each electron $i$ in its molecular state is also estimated using the relation:
\begin{eqnarray}
	E^{mol}_{i} &=& I^{B}_i+\frac{q^{A_{in}}_i}{R^{in}_{i}}\label{I-mol-i}
\end{eqnarray}

When the projectile reaches the minimum internuclear distance corresponding to the impact parameter $||$$\vec{b}$$||$, the way-in stage is ended and the way-out stage starts. At this transition point, the maximum number of shared electrons, noted $N_B$, has been reached. On the way-out, as the projectile goes away from the target, the Coulomb barrier height increases with the internuclear separation between the projectile and the target. When the Coulomb barrier reaches the binding energy $E^{mol}_{i}$ of a shared electron $i$, this electron is then either captured by the projectile or recaptured by the target. The corresponding critical distance $R_i^{out}$ is called capture radius. We consider that each electron captured by the projectile in the way-out modifies both the projectile and the target effective charges. When dealing with an electron $i$ in the way-out, the effective charges of the projectile and of the target are then respectively $q^{A_{out}}_i=q-c_i$ and $q^{B_{out}}_i=i+c_i$, where $c_i$ is the number of electrons that were previously captured by the projectile.  The value of $R_i^{out}$ given by Eq.\ref{R_out} can thus be different from the one of $R_i^{in}$:
\begin{eqnarray}
	R^{out}_{i} &=& R^{in}_{i}\frac{(\sqrt{q^{A_{out}}_i}+\sqrt{q^{B_{out}}_i})^2}{(\sqrt{q^{A_{in}}_i}+\sqrt{q^{B_{in}}_i})^2}.\label{R_out}
\end{eqnarray}
For each crossing of the projectile with a sphere of radius $R_i^{out}$ (Fig.\ref{COBM}(a)), the capture probability of the electron $i$ by the projectile is then estimated using Eq.\ref{P_i}. This probability is simply based on the multiplicities of quantum states associated with the effective principal quantum numbers, $n_i$ and $m_i$, that can be populated by the electron $i$ in the case of a capture by the projectile or of a recapture by the target, respectively.
These effective quantum numbers $n_i$ and $m_i$ are reals and are simply estimated in the framework of the Bohr model  using Eq.\ref{n_i} and Eq.\ref{m_i}, where $m_0$ is the principal quantum number of the outer shell electron on the target. They depend on the final binding energy of each electron once captured by the projectile $E_i^{A}$ (Eq.\ref{E-proj}) or recaptured by the target $E_i^{B}$ (Eq.\ref{E-targ}).
\begin{subequations}
\begin{eqnarray}
	P_i &=& \frac{n_i^2}{m_i^2+n_i^2}\label{P_i}\\
	n_i &=& \frac{q^{A_{out}}_i}{\sqrt{2E^{A}_i}}\label{n_i}\\
	m_i &=& \frac{q^{B_{out}}_i}{\sqrt{2E^{B}_i}}-\frac{q^{B_{out}}_i}{\sqrt{2I^{B}_{i+c_i}}}+m_0.\label{m_i}\\
	E^{A}_i &=& E^{mol}_{i}-\frac{q^{B_{out}}_i}{R^{out}_i}\label{E-proj}\\
	E^{B}_i &=& E^{mol}_{i}-\frac{q^{A_{out}}_i}{R^{out}_i}.\label{E-targ}
\end{eqnarray}
\end{subequations}
To summarize, the electrons $i$ = 1 to $i$ = $N_B$ will first populate molecular states in the way-in, as the projectile crosses the $R_i^{in}$ radii. On the way-out, $i$ going from $N_B$ to 1, these electrons will be redistributed on the projectile or on the target, according to their capture probability. This redistribution occurs at a distance $R_i^{out}$, with a probability $P_i$ to be captured by the projectile and a probability (1-$P_i$) to be recaptured by the target.
\subsubsection{\label{sec2-1-1}Monte-Carlo Simulation}
The model proposed by Niehaus \cite{Niehaus86} provides capture cross sections by combining the electron capture probabilities and the ranges of impact parameter leading to all possible capture multiplicities. This approach is not anymore adapted to a more complex target comprising two sites, as in the case of a dimer target. We therefore chose to combine the COBM method with Monte Carlo (MC) simulations. The MC simulation allows to set randomly the projectile coordinates noted A($x_A$, $y_A$) in the plane perpendicular to the projectile propagation axis z. In the case of an atomic target, the target position is simply defined as B($x_B$=0, $y_B$=0, $z_B$=0). For each simulated event, the successive crossing points between the projectile trajectory and spheres of radius $R^{in}_{i}$ centered on the target are determined, up to $i=N_B$ (Fig.\ref{COBM}(a)). The same approach is used in the way-out, as the projectile crosses the different spheres of radius $R^{out}_{i}$ from $i=N_B$ to $i=1$. For each capture crossing point of the way-out, a random number (uniform distribution between 0 and 1) is compared to the capture probability given by Eq.\ref{P_i} to determine whether the electron $i$ is captured by the projectile or recaptured by the target. Integrating all the events leading to a given capture multiplicity provides then directly the associated cross section. As it will be shown in the following, the use of the MC simulation greatly simplifies the treatment of molecular targets and provides better insight of specific processes through the study of the corresponding 2D probability map in impact parameter \cite{Iskandar14}. In addition, this method gives easy access to differential cross sections in transverse momentum exchange.
\subsubsection{\label{sec2-1-2}Transverse Momentum Exchange Calculation}
To get further information about the collision dynamics, the transverse momentum exchange due to the Coulomb repulsion between the projectile and the target along the collision is calculated. This information is directly connected to the projectile scattering angle which provides additional points of comparison with experimental data.

\begin{figure}
\includegraphics[width = 1.\columnwidth]{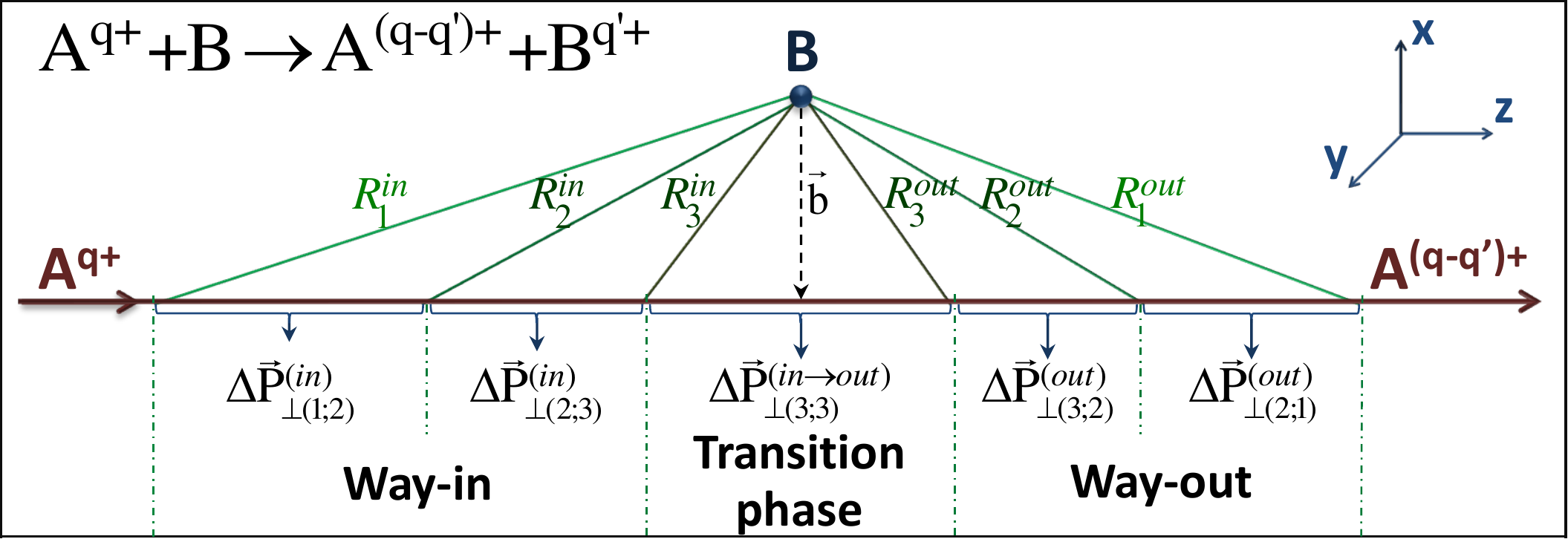}
\caption{Schematic view of the interaction between a projectile A$^{q+}$ and a target B with $N_B$ electrons involved in the way-in ($N_B$=3 in this example) and where 0 to $N_B$ electrons can be captured by the projectile in the way-out. The transverse momentum exchange, due to a Coulomb repulsion along the interaction between the projectile and the target, is calculated for all successive steps in each of the three (way-in, transition step and way-out) stages of the collision.}
\label{COBM_2}
\end{figure}

Our model of momentum transfer is based on the partitioning of the collision into a series of charge exchange. The projectile trajectory is thus segmented into steps, defined by the successive sharing and/or capture radii, with their corresponding projectile and target effective charges, $q^{A}_i$ and $q^{B}_i$. The transverse momentum exchanges calculated for each step are then cumulated to obtain the total momentum exchange of the collision. We still assume projectile straight line propagation (in the z direction), which is an acceptable approximation in the range of velocities considered here. The projectile positions at crossing points with the different sharing and capture radii, noted $z^{in}_i$ in the way-in and $z^{out}_i$ in the way-out, are calculated using the corresponding sharing radii $R^{in}_i$ or capture radii $R^{out}_i$ and the distance between the two collision partners in the transverse plane:
\begin{eqnarray}
	z^{in}_i & = z_B - \sqrt{(R^{in}_{i})^2-(x_A-x_B)^2-(y_A-y_B)^2}\nonumber\\
 	z^{out}_i &= z_B+\sqrt{(R^{out}_{i})^2-(x_A-x_B)^2-(y_A-y_B)^2}\label{z-in-out}
\end{eqnarray}

As the projectile and the target start to share electrons, the two partners of the collision exert on each other a repulsive force due to their effective charge.
The estimation of the resulting transverse momentum exchange vector all along the collision is divided in three stages: the way-in, from $z^{in}_{1}$ to $z^{in}_{N_B}$, the way-out, from $z^{out}_{N_B}$ to $z\rightarrow \infty$, and a transition stage (Fig.\ref{COBM_2}) between the way-in and the way-out introduced for convenience.

In the way-in, for each shared electron $i$ the transverse momentum exchange is integrated along the corresponding step using Eq.\ref{P-in}, from $t_i^{in}$, the time of the molecularization of the electron $i$, to $t_{i+1}^{in}$, the time of the molecularization of the next electron $(i+1)$. This calculation takes into account the effective charge of the projectile $q^{A_{in}}_i$ and of the target $q^{B_{in}}_i$ as well as the distance between the two centers in the transverse plane (corresponding to the modulus of the impact parameter $\vec{b}$($x_A-x_B$; $y_A-y_B$) for ion-atom collisions). The times $t_i^{in}$ and $t_{i+1}^{in}$ are determined using the initial velocity of the projectile $v_{p}$ (propagating in the z direction) and its z coordinates, $z_i^{in}$ and $z_{i+1}^{in}$, given by Eq.\ref{z-in-out}. 

\begin{subequations}
\begin{align}
	\Delta{\vec{P}^{(in)}_{\perp(i;i+1)}} &= \int_{t_i^{in}}^{t_{i+1}^{in}}dt\frac{q^{A_{in}}_i q^{B_{in}}_i}{(b^2+(v_{p}*t)^2)^{\frac{3}{2}}}\vec{b}\nonumber\\
		& = \frac{q^{A_{in}}_i q^{B_{in}}_i }{v_p b^2 } \left( F(t^{in}_{i+1}) - F(t^{in}_i) \right) \vec{b}\label{P-in}
\end{align}
with
\begin{align}
	F(t) & = \frac{t}{\sqrt{\frac{b^2}{v_p^2}+t^2}}\nonumber\\
	t_i^{in} &= \frac{z^{in}_{i}}{v_{p}} \ \ \text{and} \ \ t_{i+1}^{in} = \frac{z^{in}_{i+1}}{v_{p}}\nonumber
\end{align}

Similarly, the momentum exchange occurring during the transition stage is given by Eq.\ref{P-in-out}. This stage comprises only one step, the time period between $t_{N_B}^{in}$ and $t_{N_B}^{out}$, corresponding respectively to the molecularization and capture treatments of the last shared electron numbered $N_B$.
\begin{align}
	\Delta{\vec{P}^{(in \rightarrow out)}_{\perp(N_B;N_B)}} &= \frac{q^{A_{in}}_i q^{B_{in}}_i }{v_p b^2 } \left( F(t^{out}_{N_B}) - F(t^{in}_{N_B}) \right) \vec{b}\label{P-in-out}\\
	\text{with}\ \ t_{N_B}^{in} &= \frac{z^{in}_{N_B}}{v_{p}} \ \ \text{and} \ \ t_{N_B}^{out} = \frac{z^{out}_{N_B}}{v_{p}}\nonumber
\end{align}

On the way-out, the transverse momentum exchange is calculated using Eq.\ref{P-out} for the $N_B$ steps related to each capture or recapture of the electron $i$. The last step extends from $t_1^{out}$ to $t_{0}^{out}\rightarrow \infty$, as both the projectile and target may be charged after the collision.
\begin{align}
	\Delta{\vec{P}^{(out)}_{\perp(i;i-1)}} &= \frac{q^{A_{out}}_{i-1} q^{B_{out}}_{i-1} }{v_p b^2 } \left( F(t^{out}_{i-1}) - F(t^{in}_{i}) \right) \vec{b}\label{P-out}\\
	\text{with}\ \ t_{i}^{out} &= \frac{z^{out}_{i}}{v_{p}} \ \ \text{,} \ \ t_{i-1}^{out} = \frac{z^{out}_{i-1}}{v_{p}}\ \ \text{and} \ \ t_0^{out}\rightarrow\infty\nonumber
\end{align}
\end{subequations}

The total transverse momentum exchange vector $\Delta{\vec{P}_{\perp}}$ is obtained by summing the ($N_B$-1) steps of the way-in, the step of the transition stage and the $N_B$ steps of the way-out:
\begin{align}
	\Delta{\vec{P}_{\perp}} &= \sum^{N_B-1}_{i=1}\Delta{\vec{P}^{(in)}_{\perp(i;i+1)}}+\Delta{\vec{P}^{(in \rightarrow out)}_{\perp(N_B;N_B)}}+\sum^{1}_{i=N_B}\Delta{\vec{P}^{(out)}_{\perp(i;i-1)}}\label{P-tot}
\end{align}

The effective charges of the projectile $q^{A_{in(out)}}_i$ and of the target $q^{B_{in(out)}}_i$ used in Eq.\ref{P-in}, \ref{P-in-out} and \ref{P-out} depend on the number of molecular electrons in the way-in and in the transition phase. In the way-out, they depend on both, the number of molecular electrons $i$ and the number of captured electrons $c_i$. As in the model of Niehaus, the charge of an electron captured by the projectile or recaptured by the target in the way-out will be accounted for in the effective charge of its new center (projectile or target). In a classical approach, this choice can be easily justified by the fact that the classical orbital radius of the captured or recaptured electron is then smaller than the internuclear distance between projectile and target. It is more difficult to deal with the electrons that are shared by both centers after their molecularization in the way-in. In the COBM cross sections calculation as proposed by Niehaus, shared electrons are pure spectators until capture by the projectile or recapture by the target (Sec.\ref{sec2-1-0}). But for the calculation of the transverse momentum exchange, the partial screening effect of the $i$ shared electrons on the target effective charge might play an important role. To take into account the effective distribution of the electrons around the molecule in each segment of the trajectory, the screening could be introduced as a position-dependent function in integrals \ref{P-in}, \ref{P-in-out} and \ref{P-out}, as it is a common practice in nuclear stopping theories. For the sake of simplicity and to limit as much as possible the number of free parameters, we introduce in the calculation of the transverse momentum transfer a single screening parameter, noted $S$, whose effect only depends on the number of shared electron $i$ in each piece of the trajectory. The value of $S$ can be adjusted between 0 and 1 according to different screening scenarios (see Sec. \ref{sec4-2}). The effective charges of the projectile and of the target are then respectively given by:
\begin{subequations}
\begin{equation}
	q^{A_{in}}_i = q  \ \ \text{and} \ \ q^{B_{in}}_i =i \times (1-S)\label{qeff-in}
\end{equation}
in the way-in (and in the transition phase with $i=N_B$) and by
\begin{equation}
	q^{A_{out}}_i = q -c_i \ \ \text{and} \ \ q^{B_{out}}_i = i\times (1-S)+c_i\label{qeff-out}
\end{equation}
\end{subequations}
in the way-out.
Note that the screening parameter $S$ is only introduced into the model for the calculation of transverse momentum exchange, while capture cross section calculations simply use the effective charges given in Sec.\ref{sec2-1-0}.
\subsection{\label{sec2-2}Collisions with Atomic Dimer Targets}
The MC-COBM model can be applied to dimer targets when making the approximation that the dimer can be treated as two independent atoms fixed in space. This approximation is justified by the low electron mobility between the two atoms of the dimer and by its large internuclear distance. The same approach as for a simple atomic target can then be used, by considering the collision of the projectile with the two centers B and C of the dimer as two separate collisions with two atoms. The only indirect effect of one site on the other is the change of the projectile effective charge during the collision, as this change is then caused by captured electrons from both centers of the dimer. This indirect influence of a second atomic center combined with the increase of the degrees of freedom due to the geometry of diatomic molecular target are the major motivations for using MC simulations.
\begin{figure}
\includegraphics[width=1.\columnwidth,clip]{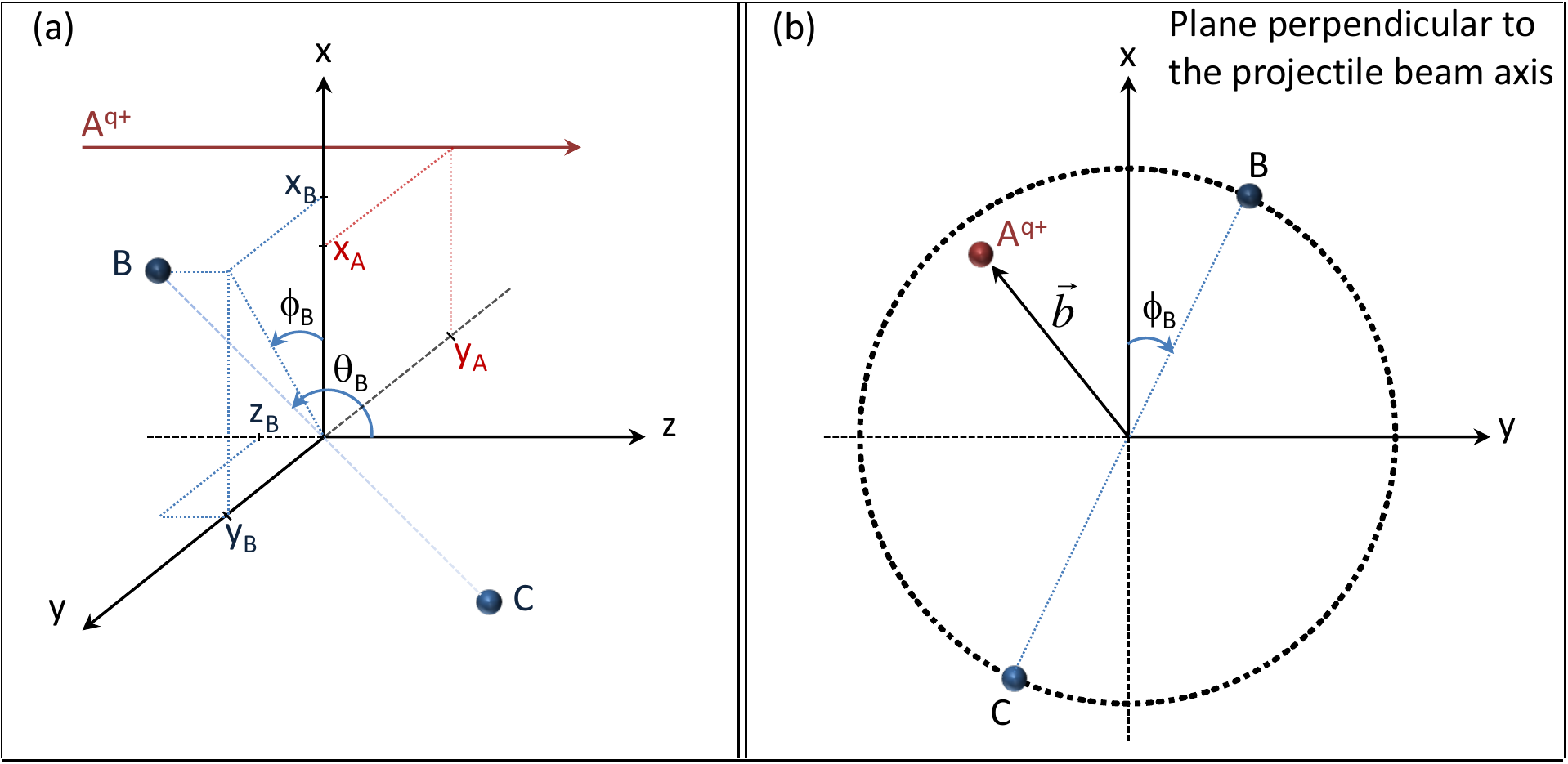}
\caption{Representation of the collision between a projectile A$^{q+}$ and a dimer target B-C in a 3D Cartesian coordinate system (x, y, z) (a) and in the transverse plan (x, y) (b). ($x_B$, $y_B$, $z_B$) are the coordinates of the center B of the molecule, $\theta_B$ is the angle of the molecular axis with respect to the projectile beam axis (z axis), $\phi_B$ is the projected angle of the molecular axis in the transverse plane with respect to x axis and ($x_A$, $y_A$) are the coordinates of the projectile in the transverse plane.}
\label{geometry}
\end{figure}
All the possible trajectories of the projectile with respect to the positions of the two centers of the dimer (Fig.\ref{geometry}) are taken into account by generating, in a random way, the position ($x_A$, $y_A$) of the projectile in the transverse plane and the orientation of the dimer around its center of gravity arbitrarily fixed at (0, 0, 0) coordinates. The orientation of the dimer is obtained by sampling randomly the two parameters cos($\theta$) and $\phi$, where $\theta$ is the polar angle of the dimer axis with respect to the z axis and $\phi$ is the angle of the dimer axis in the transverse plane (x, y) with respect to the x axis. For the internuclear distance between centers B and C, we used the values 5.86 a.u. and 7.18 a.u. for Ne$_2$ and Ar$_2$, respectively.  As the collision duration is very short compared to the vibrational period of the dimer target, these distances are considered as constant.  
\subsubsection{\label{sec2-2-1}Capture cross sections calculations}
The COBM model is then applied as described in Sec.\ref{sec2-1} for atomic targets, but considering two atomic centers whose electrons can be shared and captured separately.
In a first step, the number of electrons potentially occupying a molecular state in the way-in, noted $N_B$ for the center B and $N_C$ for the center C of the dimer, are determined. These shared electrons are now numbered $i_B$ and $i_C$, depending on their center of origin. The associated sharing radii and the binding energy of electrons $i_B$ and $i_C$ are calculated using Eq. \ref{R_in} and \ref{I-mol-i}. As the projectile enters the way-out of one of the two centers, the capture by the projectile or the recapture by the target of the molecular electrons may occur accordingly to Eq.\ref{P_i} at the corresponding capture radii. Note that within this model, the electrons are shared between the projectile and their center of origin, but not between the two centers of the target. Shared electrons can thus only be recaptured on their initial atomic site or captured by the projectile.
As mentioned previously, the possible change in charge of the projectile is taken into account to determine the next crossing radii with the two centers, for capture or molecularization. When treating the electrons of center B using Eq.\ref{R_in} to Eq.\ref{E-targ}, $i$ is thus replaced by $i_B$ and the target effective charge is simply $q^{B_{in}}_{i_B}=i_B$ on the way-in. When dealing with an electron $i_B$, we now have to account for the number of electrons previously captured by the projectile from targets B and C, still noted respectively $c_{i_B}$ and $c_{i_C}$. The target B effective charge thus becomes $q^{B_{out}}_{i_B}={i_B}+c_{i_B}$ on the way out, and the projectile effective charge $q^{A_{in}}_{i_B}=q-c_{i_C}$ on the way-in and $q^{A_{out}}_{i_B}=q-c_{i_B}-c_{i_C}$ on the way-out. The same formulas apply to electrons $i_C$ from the target C by swapping the subscripts or superscripts $B$ and $C$. Because of the different possible orientations of the dimer, molecularization and capture of electrons from B and C can occur in different order. This implies that the evaluation of the crossing radii and the determination of capture or recapture of the electrons have to be done sequentially to know at anytime $i_B$, $i_C$, $c_{i_B}$ and $c_{i_C}$, as the projectile progresses along the collision axis.
At the end of the collision, we keep track of the number of electrons captured from each center of the dimer. The corresponding charge distribution on the ionized dimer is associated to a capture channel noted ($q_B$, $q_C$)$_c$ where $q_B$ and $q_C$ are the charge states of the center B and C after the collision. In addition, the number $i_B$ and $i_C$ of the individual captured electrons are recorded, giving access to their initial atomic shell \cite{Iskandar15}, as well as the initial coordinates of the projectile and of the dimer sites B and C for the study of impact parameter dependence \cite{Iskandar14}.
\subsubsection{\label{sec2-2-2}Transverse momentum calculations}
The transverse momentum exchange along the collision is also calculated separately for the center B and for the center C, using the methodology described in Sec.\ref{sec2-1-2}. Compared to the atomic case, we now have to account for a projectile effective charge including the contributions from shared and captured electrons from both centers of the dimer. This implies a segmentation of the projectile trajectory in as many steps as defined by the total number of crossing radii involved when considering both centers. When dealing with an electron $i_B$ from the center B, the projectile and target effective charges are thus defined as 
\begin{subequations}
\begin{equation}
	q^{A_{in}}_{i_B} = q -c_{i_C}\ \ \text{and} \ \ q^{B_{in}}_{i_B} =i_B\times (1-S)\label{qeff-in-dimer}
\end{equation}
in the way-in (and in the transition phase with $i_B=N_B$) and by
\begin{eqnarray}
	q^{A_{out}}_{i_B} = q -(c_{i_B}+c_{i_C}) \nonumber\\
	 \text{and} \ \ q^{B_{out}}_{i_B} = i_B\times (1-S)+c_{i_B}\label{qeff-out-dimer}
\end{eqnarray}
\end{subequations}
in the way-out, where $S$ is the partial screening parameter introduced in Sec. \ref{sec2-1-2}. Again, the same formulas apply to electrons from the target C by swapping the subscripts or superscripts $B$ and $C$. 
The sum of the contributions from both sites provides the total transverse momentum exchange induced by the collision between the projectile and the dimer.
\section{\label{sec3}Post-Collision Treatment of the MC-COBM data}
For a meaningful comparison of the MC-COBM results with those obtained experimentally, several additional steps are required. First, one has to determine the fragmentation channels expected for each capture channel provided by the model. Then, the trajectories of the two ionic fragments from the dimer and the response function of the experimental setup have to be simulated.
\subsection{\label{sec3-1}Estimation of Relaxation Channels}
The MC-COBM allows us to determine the number of electrons captured by the projectile from each center of the dimer and thus, to deduce the charge distribution on the dimer ($q_B$;$q_C$)$_c$ for each simulated event right after the collision. A first step before any comparison with the experimental results consists in identifying the relaxation channels leading to the fragmentation of the molecular ion. The resulting fragmentation channels are noted ($q_1$;$q_2$)$_f$, where the subscript $f$ stands here for fragmentation, unlike the subscript $c$, indicating a capture channel. This step is essential as the charge states $q_1$ and $q_2$ may not be the same as $q_B$ and $q_C$, and because the experiment is sensitive to charged fragments only.

Several scenarios can be considered for the relaxation mechanisms of a given capture channel ($q_B$;$q_C$)$_c$ towards fragmentation channels ($q_1$;$q_2$)$_f$. One of these scenarios corresponds to the capture of electrons from both centers of the dimer, leaving the dimer in a dissociative state which directly relax in two charged fragments. Here, the dissociation occurs directly via Coulomb Explosion (CE) and the charge distribution on the dimer remains the same (($q_B$;$q_C$)$_c$ $\equiv$ ($q_1$;$q_2$)$_f$) and 
other scenarios have to be considered when the projectile captures electrons from only one center of the dimer, leaving the dimer in an asymmetric charge distribution with one charged center and one neutral center. Here, the collision can either populate a dissociative state or a non-dissociative state of a specific channel. When a dissociative state of the dimer with one neutral site is populated, the simulated event is excluded from the comparison with the experimental data as neutral fragments are not detected. As it will be discussed in the following sections, the determination of this probability to populate unbound states with one neutral center is non trivial.

For the non-dissociative states, the two centers of the dimer remain bound until relaxation towards a dissociative state via intermediate mechanisms such as Charge Transfer (CT),  Radiative Charge Transfer (RCT) or Interatomic Coulombic Decay (ICD) \cite{Matsumoto10,Iskandar15}. These processes lead, respectively by charge exchange and energy transfer between the two centers of the dimer, to the production of two charged fragments to which the experimental detection system is sensitive. The relaxation process of a given capture channel is determined after the study of the molecular ion potential energy curves (PECs), as previously discussed in~\cite{Matsumoto10,Iskandar15}. In most cases, direct crossings between PECs of molecular states associated with a channel ($q_B$;$0$)$_c$ and molecular states associated with ($q_B$-1;1)$^{*}_{c}$  lead to the ($q_B$-1;1)$_{f}$ fragmentation channel through CT. In the present collision systems, this is the case for the (4;0)$_c$ and (3;0)$_c$ capture channels. As shown in~\cite{Matsumoto10,Iskandar15,Durand92}, such direct crossings do not exist for the one-site double capture channel (2;0)$_c$ and the latter relaxes towards the (1;1)$_f$ fragmentation channel through RCT.

For single capture (1;0)$_c$, the non-dissociative states are stable and do not dissociate, except for the case of inner valence shell single electron capture from a Ne dimer. In that specific case, the excitation energy is sufficient for the ICD process to occur, resulting in the population of the (1;1)$_f$ fragmentation channel with the emission of a low energy electron, as previously observed in \cite{Iskandar15}. With Ar$^{9+}$ and Xe$^{20+}$ projectiles, the experimental results have shown no sign of ICD and the MC-COBM does not predict any significant inner-shell single electron capture from Ne dimers. The ICD process will thus not be discussed further in the present article.

After simulating many events (typically $10^8$ per collision system), the MC-COBM data are sorted according to the relaxation process and to the final fragmentation channel inferred using the method described above. This classification is kept in the following steps of the simulation, providing the relative production of the different relaxation mechanisms and fragmentation channels (Sec.\ref{sec4-1}) and the associated transverse momentum exchange distributions (Sec.\ref{sec4-2}).
\subsection{\label{sec3-2}Simulation of the Spectrometer and Analysis of the Data}
This section describes briefly the principle of the experimental setup and how the trajectories of the charged fragments inside the spectrometer are simulated. The full details of the COLTRIMS experimental apparatus are described in Refs. \cite{Matsumoto10,Matsumoto11,Iskandar14,Iskandar15}. The projectile ion beam crosses the dimer target provided by a supersonic gas jet in the center of the recoil ion momentum spectrometer. After the collision, recoil ions and dimer fragments resulting from charge transfer are collected using the uniform electric field of the spectrometer and detected by a microchannel plates delay-lines detector giving both the detection time and position of detected ions. The coincidence TOF (time of flight) map of both fragments from a dimer dissociation is used to identify and select the fragmentation channels. The fragments momenta are calculated from the positions and TOF data by imposing momentum conservation restriction for optimal resolution and false coincidence events suppression \cite{Matsumoto11}. The Kinetic Energy Release (KER), deduced from the fragments momenta in the molecular frame, provides then an identification of the relaxation mechanisms leading to a specific fragmentation channel. By separating all the processes, the relative intensities of all relaxation mechanisms is finally determined. In addition, as shown in \cite{Iskandar14}, the momenta of the two fragments give access to the initial orientation of the dimer target and to the transverse momentum exchange induced by the collision.

The geometry of the spectrometer, the size of the collision region, and the electric field applied to extract the fragments all the way to the detector have been implemented in last step of the MC-COBM simulation. This final treatment of the MC-COBM events ensures an unambiguous and fair comparison between experiment and theory.  
The spatial extension of the collision region is simulated by generating randomly the initial position of the dimer using a 3D Gaussian distribution of FWHM 0.6 mm corresponding to the overlap between the projectile beam and the gas jet. Each process leading to a given fragmentation channel releases a specific kinetic energy for the two charged fragments \cite{Matsumoto10,Iskandar15}. For dimers, this KER depends, to first order, on the charge state of the two fragments and on the internuclear distance at which the dissociation takes place. In the simulation, we have used the distributions in KER obtained experimentally for each process leading to a specific fragmentation channel: for each simulated event, the KER is randomly selected according to the proper experimental distribution. The momentum vector of the fragments in the frame of the dimer center of mass is deduced from the KER and from the orientation of the dimer.
The momentum exchange between the projectile and the target provided by the MC-COBM calculation is then accounted for to determine the momentum vector of each fragment in the laboratory frame.  For each event, the trajectory of the two charged fragments inside the extraction field of the spectrometer is calculated using these initial parameters: position, momentum, mass and charge of the fragments. The TOF and the position on the detector of the two ions are finally convoluted with the detector response function (position and TOF resolution of respectively 0.5~mm and 0.5~ns FWHM)  and recorded. 

Following the spectrometer simulation, the last step consists in analyzing the simulated data using the same procedure as for the experimental data. This method ensures that all the relevant apparatus effects are included in the simulations prior confrontation with the experimental results.
\section{\label{sec4}MC-COBM vs Experimental Results}
Part of our experimental results on collisions between MCIs projectiles and rare gas dimers have already been presented in previous publications \cite{Matsumoto10,Iskandar14,Iskandar15}.
The results of Matsumoto et al. \cite{Matsumoto10} were obtained using Ar$^{9+}$ projectiles colliding on Ar$_2$ targets. They have shown evidence for a large charge asymmetry in the dissociation of Ar dimers following multiple-electron capture, as opposed to what was previously seen using covalent molecules such as N$_2$ \cite{Itzhak93,Ehrich02}. Moreover, for the fragmentation channel (1;1)$_f$, an unambiguous separation between two-site capture, related to CE process, and one-site capture, related to RCT process, was observed using the KER distribution. Later on, a comparison between the MC-COBM model and these experimental results was performed \cite{Iskandar14}. A very good agreement has been obtained in terms of relative production of the different dissociation mechanisms and associated fragmentation channels.

For the double electron capture channel, the relative contributions of CE, RCT, and ICD given by the calculations for three different collision systems involving O$^{3+}$, Ar$^{9+}$ and Xe$^{20+}$ projectiles and Ne$_2$ targets were also compared to the experimental data \cite{Iskandar15}. The calculations reproduced reasonably well the CE versus RCT contributions for both Ar$^{9+}$ and Xe$^{20+}$ projectiles, with a similar dependence of these contributions on the projectile charge state. Moreover, the appearance of the ICD process observed experimentally in neon dimers was also predicted by the model for the low charge O$^{3+}$ projectiles. However, for this collisional system O$^{3+}+ \text{Ne}_2$, the estimation by the model of the relative production of ICD, CE and RCT processes did not reproduce quantitatively the experimental results. This indicates some of the limits of this model for low projectile charge states with a hydrogenic approximation which becomes rudimentary for a precise estimate of the capture probabilities in the way-out of the collision (Eq.\ref{P_i}).

Measurements of the angular correlation between the scattered projectile and the recoiling fragments combined with model calculations have also been investigated for the Ar$^{9+}$ + Ar$_2$ collision system. This study provided access to atomic site sensitivity, showing that electron capture from “near-site” atoms is strongly favored \cite{Iskandar14}, as opposed to what was previously observed with N$_2$ covalent molecules \cite{Ehrich02}.

To provide a more complete and stringent test of the MC-COBM approach, we present in the following an exhaustive comparison of experimental data with calculations for four collision systems involving Ar$^{9+}$ and Xe$^{20+}$ projectiles at 15~qkeV colliding with Ar$_2$ and Ne$_2$ targets. 
\subsection{\label{sec4-1}Capture and Fragmentation Channels}
For want of anything better, we simply considered in previous calculations~\cite{Matsumoto10,Iskandar14,Iskandar15} that 50~\% of the one-site electron capture yield resulted directly in the population of unbound states with the emission of one neutral fragment. There are two physical mechanisms which support this assumption.

 The first mechanism is the intersystem crossing, which couples some vibrational bound states with the continuum states of another spin multiplicity by means of spin-orbit interaction. For the X$_2^{2+}$ molecules with X=Ar,Ne, the singlet states correlating to the asymptotic channels X$^{2+}$(1D)-X and  X$^{2+}$(1S)-X are energetically above the triplet dissociation limit X$^{2+}$(3P)-X. The intersystem crossing will thus takes place, in competition with the RCT relaxation mechanism. As a radiative transition, the latter is quite slow. The RCT lifetime for Ar$_2^{2+}$ is typically 5.0 ns \cite{Daskalopoulou93} and there is sufficient time for the intersystem crossing to take place efficiently. This is confirmed indirectly by the observation of radiative emission, for which no RCT transition is observed at the energy corresponding to the singlet states for Ar$_2^{2+}$ \cite{Durand92}.

Dissociation of Ar$^{2+}$-Ar and Ne$^{2+}$-Ne states occurs directly if the vibrational energy exceeds the potential well depth of the state populated by the collision. By considering that vibrational energy can be provided to the target through momentum exchange with the projectile and that transverse momentum exchange dominates over longitudinal momentum exchange, the maximum vibrational energy brought to the target can be estimated. For collisions leading to one-site multiple capture such as the (2;0)$_c$ capture channel, the measurement of the transverse momentum exchange gives thus new insight on the probability to produce unbound states with neutral emission. As it is shown in the next section, for (2;0)$_c$ capture channels, the maximum transverse momentum exchange observed experimentally is $\sim$20~a.u.. It corresponds to a maximum energy exchange of $\sim$75~meV for Ar$_2$ targets and of $\sim$150~meV  for Ne$_2$ targets. On the other hand, the binding energy in the Franck Condon region (at internuclear distances of R=7.18 a.u. and R=5.86 a.u. for Ar$_2$ and Ne$_2$ targets, respectively) can be approximated by accounting for the polarizability term in $-\alpha q^2/2R^4$, as shown in \cite{Durand92}. We find binding energies of 236~meV for the Ar$^{2+}$-Ar states and of 123~meV for the Ne$^{2+}$-Ne states. This is in both cases too high to lead to an efficient dissociation. As the binding energy evolves with the square of the charge, a similar situation is expected for the (3;0)$_c$ and (4;0)$_c$ capture channels. A more elaborate calculation should be performed to provide a better estimate, but this simple and classical comparison indicates a possible small contribution for Ne$_2$ and a negligible amount in the case of Ar$_2$. This small amount of dissociation adds to the singlet dissociation discussed above, yielding a total dissociation fraction that should not exceed 50\%. .

\begin{figure*}
\subfloat{\includegraphics[width =1\columnwidth]{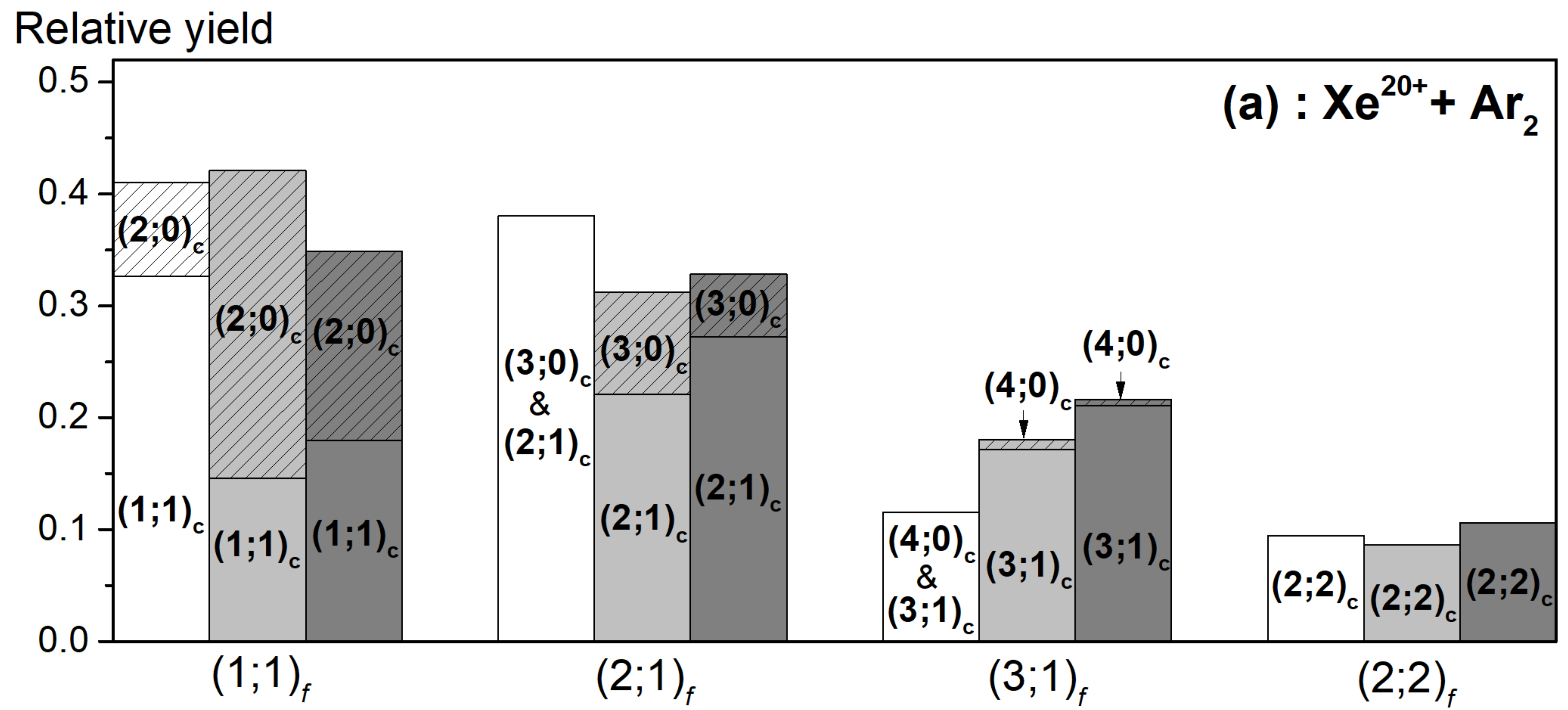}}
\subfloat{\includegraphics[width = 1\columnwidth]{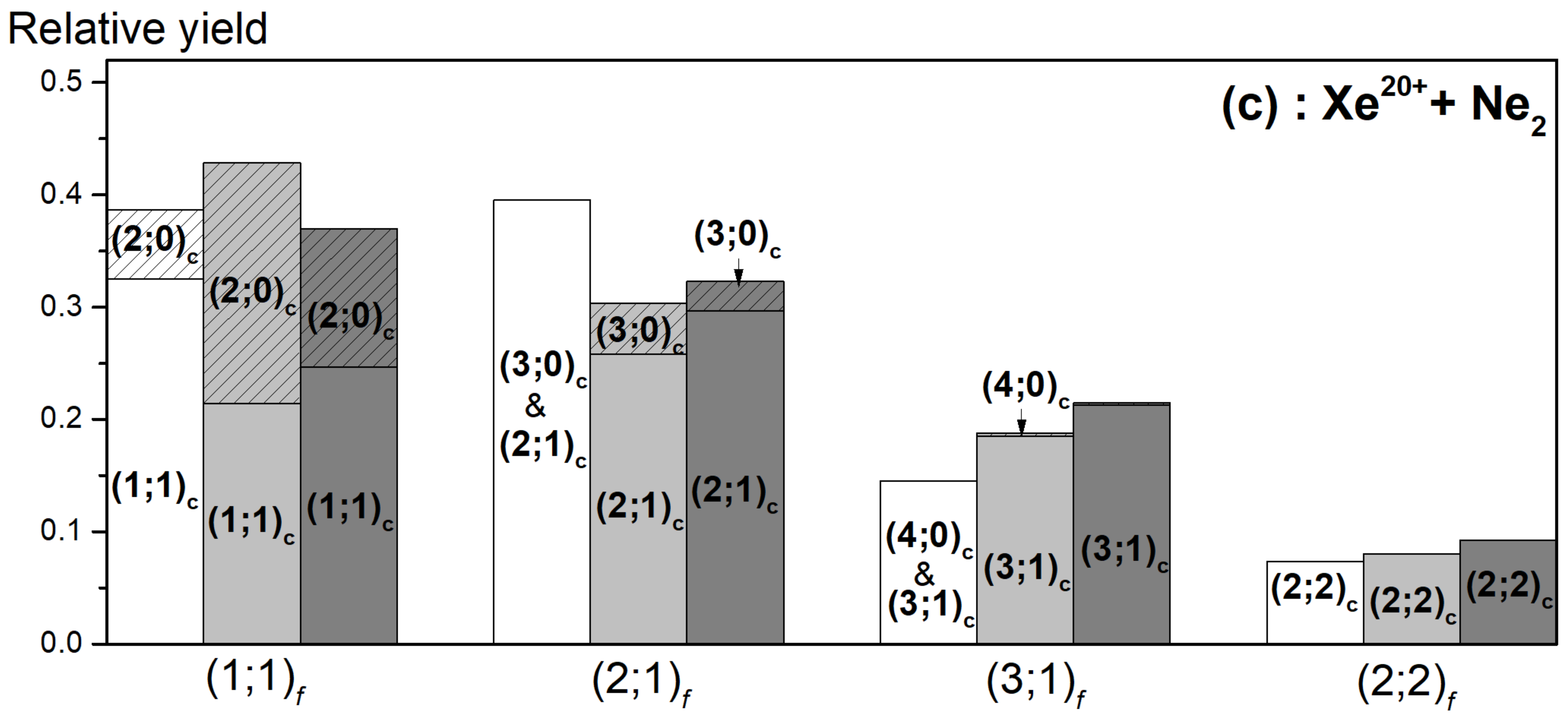}}\\
\subfloat{\includegraphics[width = 1\columnwidth]{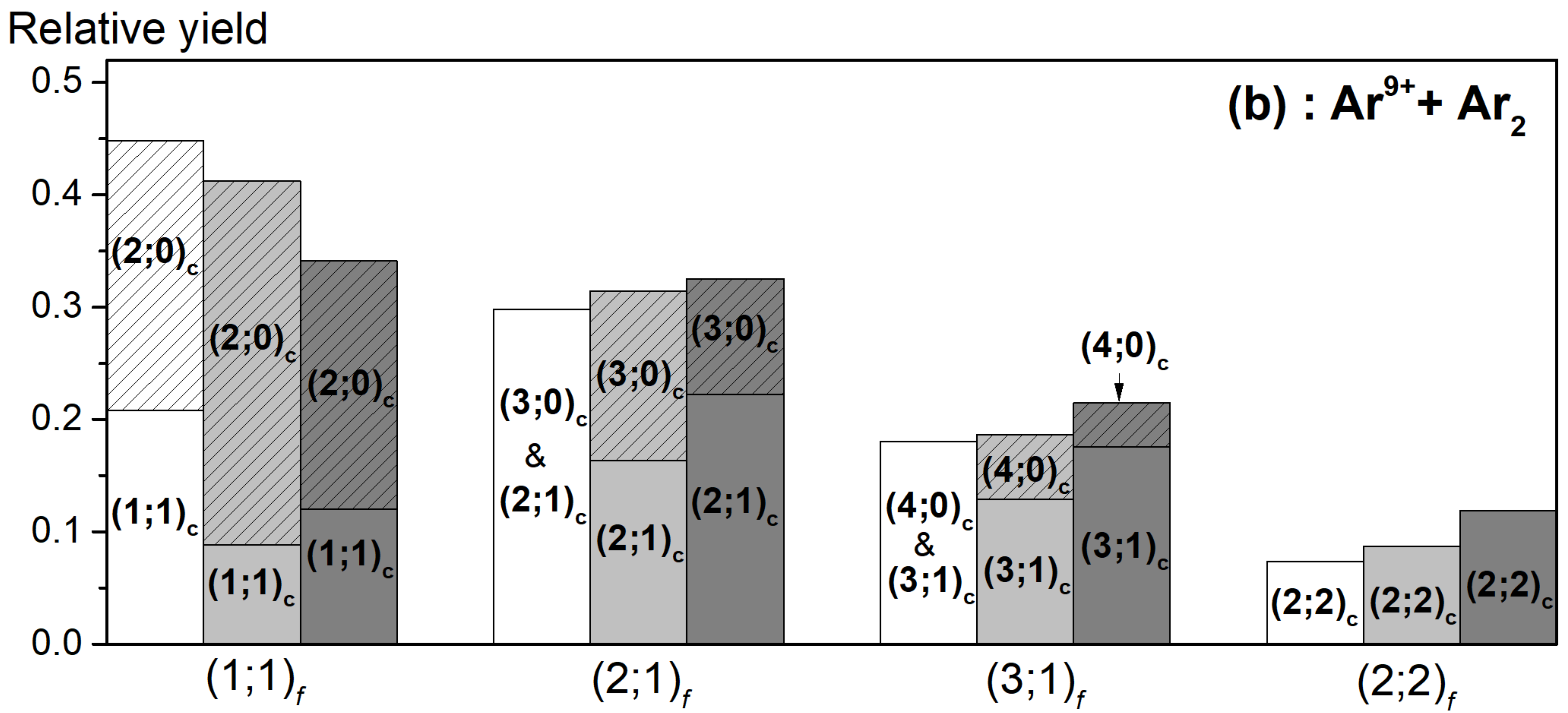}}
\subfloat{\includegraphics[width = 1\columnwidth]{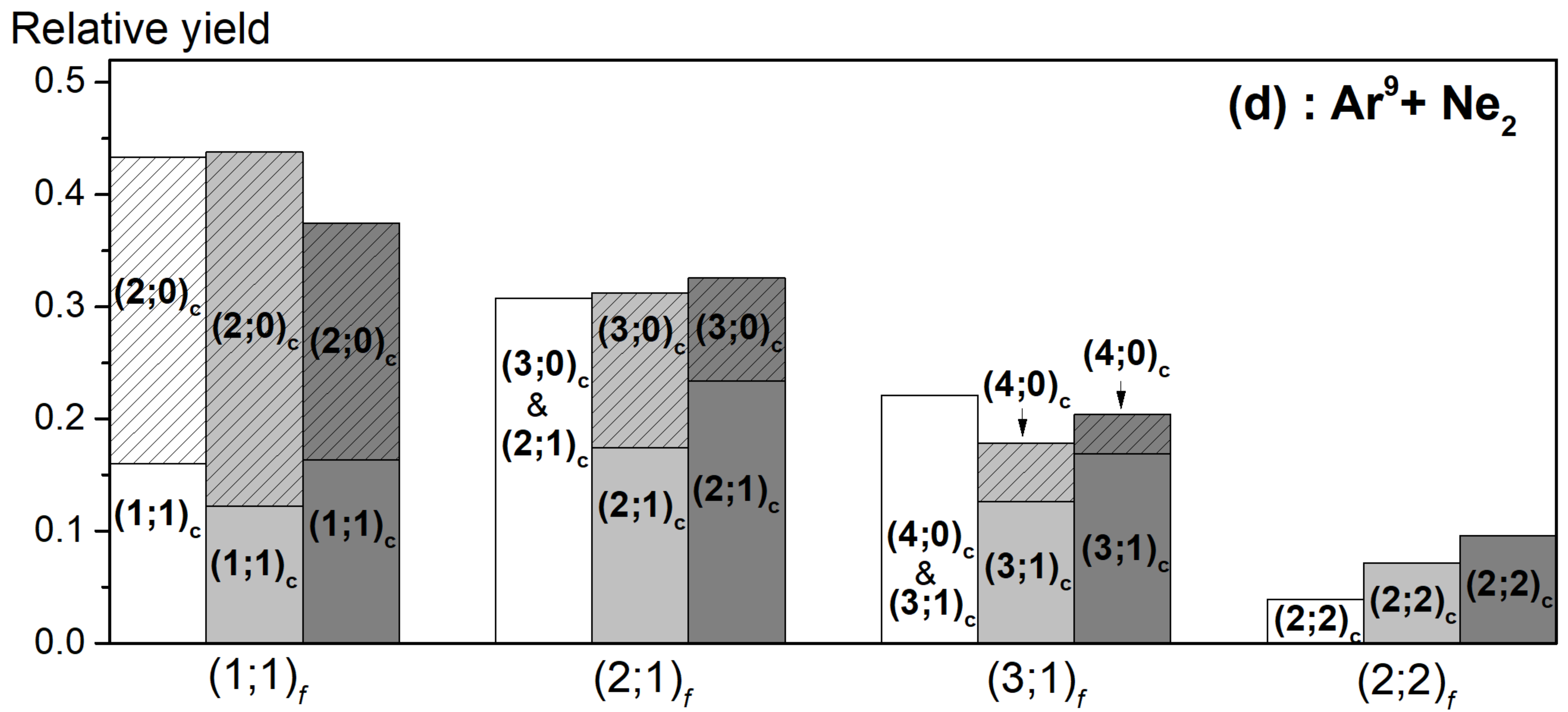}}
\caption{Relative yield of the different fragmentation channels obtained experimentally (white) and using the MC-COBM calculations. MC-COBM results are given for two different scenarios: when considering no direct production of dissociative states with one neutral fragment (light gray) and when considering equiprobable populations of dissociative states and transient non-dissociative states (gray). The contributions of the initial capture channels are indicated when relevant.}
\label{Ar-Ne-ratio}
\end{figure*}

 In the MC-COBM calculation, we thus now consider two different scenarios for the population of dissociative states for capture channels involving one-site electron capture. In the first scenario, there is no direct production of dissociative states comprising one neutral fragment, meaning that 100\% of one-site multiple capture events populate transient non-dissociative molecular states, all relaxing eventually by the emission of two charged fragments. In the second one, 50~\% of the one-site electron capture yield results in direct fragmentation (q;0)$_f$ with one neutral center, to which the experimental setup is insensitive.
For all four collision systems, the relative yields of the different fragmentation channels obtained experimentally are compared in Fig.\ref{Ar-Ne-ratio} to the MC-COBM results for the two scenarios discussed above.

As discussed in Sec. III.A, each fragmentation channel ($q_1$;$q_2$)$_f$ can be fed by different relaxation mechanisms ($q_B$;$q_C$)$_c$ corresponding to a given number of electrons removed from each site of the dimer during the collision, prior to any possible charge redistribution. The (1;1)$_f$ fragmentation channel can result from two-site Double Capture (DC) (1;1)$_c$ leading directly to Coulomb explosion (CE), but also from one-site DC (2;0)$_c$ populating non-dissociative molecular states relaxing through radiative charge transfer (RCT). These two processes can be distinguished experimentally by their different KER distribution \cite{Matsumoto10,Iskandar15} and their contributions are shown separately. Transient non-dissociative molecular states populated by one-site triple capture (TC) and quadruple capture (QC), denoted (3;0)$_c$ and (4;0)$_c$, lead to the (2;1)$_f$ and (3;1)$_f$ fragmentation channels, respectively, through charge transfer occurring at direct crossings with excited states (CT) or with radiative emission (RCT). These processes are expected to result in KER distributions higher or close to the ones obtained for the direct two-site capture channels (2;1)$_c$ and (3;1)$_c$. As shown in\cite{Matsumoto11}, both CT and RCT are here weak channels and one-site multiple capture could thus not be clearly isolated experimentally from two-site capture for TC and QC fragmentation channels.

Within the MC-COBM model, all possible capture channels leading to a given fragmentation channel are by nature separated (Sec.\ref{sec3-1}). For the (2;0)$_c$, (3;0)$_c$ and (4;0)$_c$ one-site capture channels, 100~\% (first scenario, light gray) or 50\% (second scenario, gray) of the populations given by the calculations have been attributed to transient non-dissociative molecular states. These contributions were added to the (1;1)$_f$, (2;1)$_f$ and (3;1)$_f$ fragmentation channels, fed respectively through the RCT and CT processes. For both scenarios, the MC-COBM results are found in good agreement with the experimental data of the four collision systems. The relative populations of the fragmentation channels of interest are well reproduced with maximum deviations remaining below ~10$\%$. If the  agreement is somewhat better when using the first scenario, considering that 50\% of the one-site multiple capture results in the emission of a neutral fragment does not affect strongly the final distributions shown in Fig.\ref{Ar-Ne-ratio}.

For QC channels, Fig.\ref{Ar-Ne-ratio} clearly shows the preference for the asymmetric fragmentation channels (3;1)$_f$ over the symmetric one (2;2)$_f$, specific to vdW bound atomic systems and resulting from the low electron mobility between the two atoms of the dimer. This feature is reproduced using the two scenarios.

For the DC channels, we clearly see the evolution with the charge states of the projectile of the relative contributions of the RCT and CE processes associated respectively to the (2;0)$_c$ and (1;1)$_c$ capture channels. The experimental results show that for the high charge state of the projectile Xe$^{20+}$, the CE process following a two-site double capture dominates, while for the lower charge state projectile Ar$^{9+}$, the RCT process associated to a one-site double capture takes over. This behavior can be intuitively understood using simple geometrical considerations. As shown in table \ref{capture_radii}, the sharing radii given by the COBM model for double electron capture from the 3p shell of Ar and 2p shell of Ne using Xe$^{20+}$ projectile are about two times larger than the internuclear distance of the Ar$_2$ (7.18 a.u.) and Ne$_2$ (5.86 a.u.) targets and should favor the removal of electrons from both sites of the dimer. For Ar$^{9+}$ projectiles, the double electron sharing radii are only slightly larger than the internuclear distance of the Ar$_2$ and Ne$_2$, and one-site double capture becomes more probable than two-site double capture. 

\begin{table}
\caption{Sharing radii R$_i^{in}$ (in a.u.) estimated by the COBM model, using Eq.\ref{R_in}, for double electron capture by Xe$^{20+}$ and Ar$^{9+}$  projectiles from atomic Argon and Neon targets.}
\label{capture_radii}
\begin{ruledtabular}
\begin{tabular}{ccc}
Projectile  &  Ar target  &  Ne target\\
& $R_2^{in}[3p^{-2}]$  & $R_2^{in}[2p^{-2}]$\\
\hline
Xe$^{20+}$ & 14.42 & 9.73  \\
Ar$^{9+}$ & 10.32 & 6.96  \\
\end{tabular}
\end{ruledtabular}
\end{table}
This dependence is qualitatively well reproduced by the model, with a two-site double capture contribution increasing with the projectile charge when using both scenarios. For Ar$^{9+}$ projectiles, the second scenario (50\% of unbound states, dark gray) results in one-site over two-site double capture ratios in good agreement with the experiment. But the first scenario, with no direct population of dissociative channels (2;0)$_f$, tends to overestimate one-site double capture. For Xe$^{20+}$ projectiles, the agreement worsens, as with both scenarios, one-site double capture is overestimated. These observations do not allow to conclude on the validity of one of the scenarios over the other, but indicate more likely shortcomings of the MC-COBM model when calculating the one-site and two-site double capture cross sections, in particular for Xe$^{20+}$ projectiles.
 
\subsection{\label{sec4-2}Transverse Momentum Exchange}

\begin{figure*}
\subfloat{\includegraphics[width = 1.8\columnwidth]{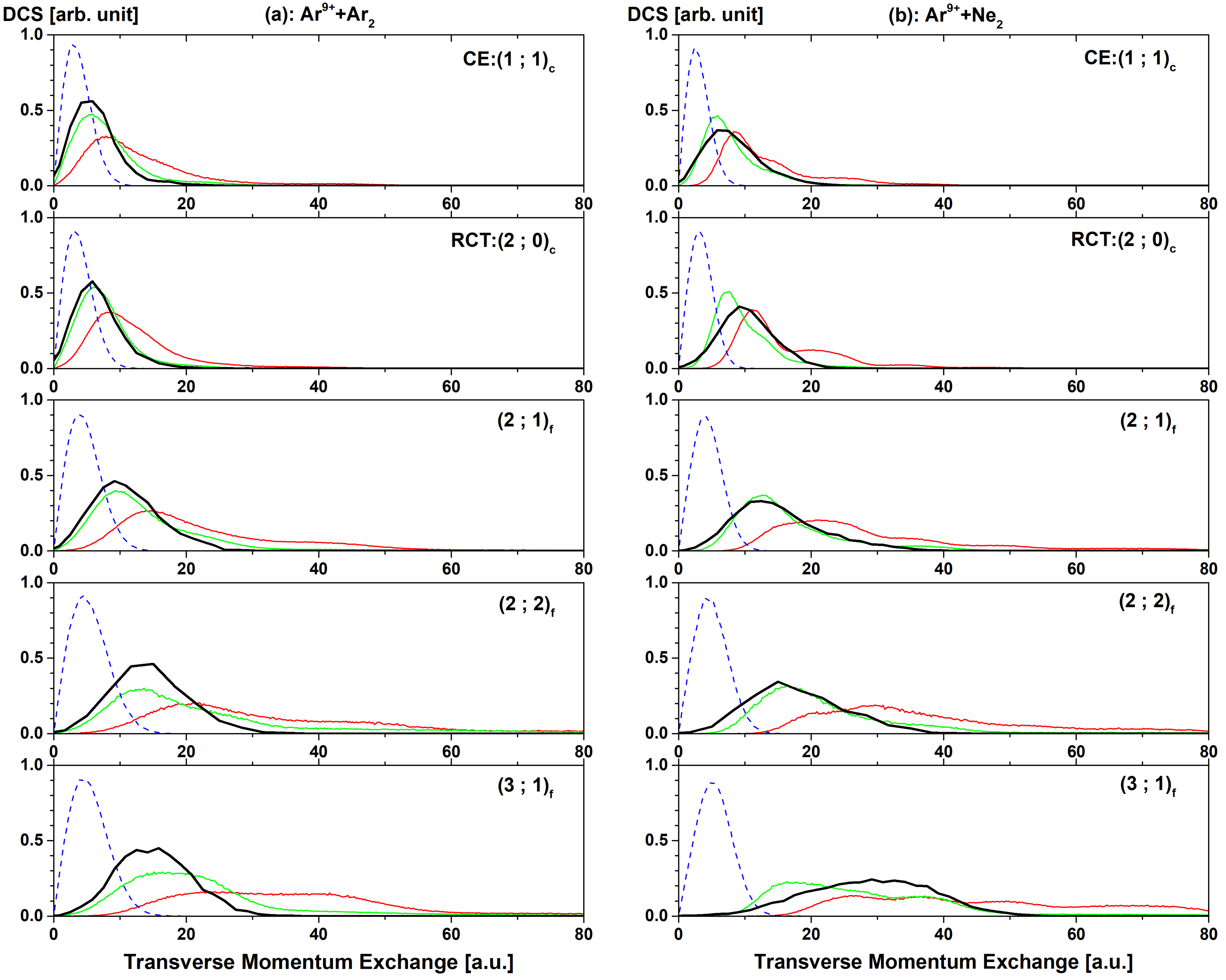}}
\caption{Differential cross section (DCS) in transverse momentum exchange obtained experimentally (thick black line) and using the MC-COBM for Ar$^{9+}$ projectiles colliding with Ar dimers (a) and Ne dimers (b). The MC-COBM distributions, previously normalized to the experimental data, are given for three different charge screening parameters: $S=1$ (blue dashed line), $S=1/2$ (green line (or gray)) and $S=0$ (red line (or dark gray)).}
\label{PCM_Ar}
\end{figure*}

The good agreement between the experimental results and the MC-COBM data shown in the last section motivated to further test the capability of the model to reproduce the experimental results. We focus in this section on the distribution of the transverse momentum exchange arising from the Coulomb repulsion between the collision partners. 

\begin{figure*}
\subfloat{\includegraphics[width = 1.8\columnwidth]{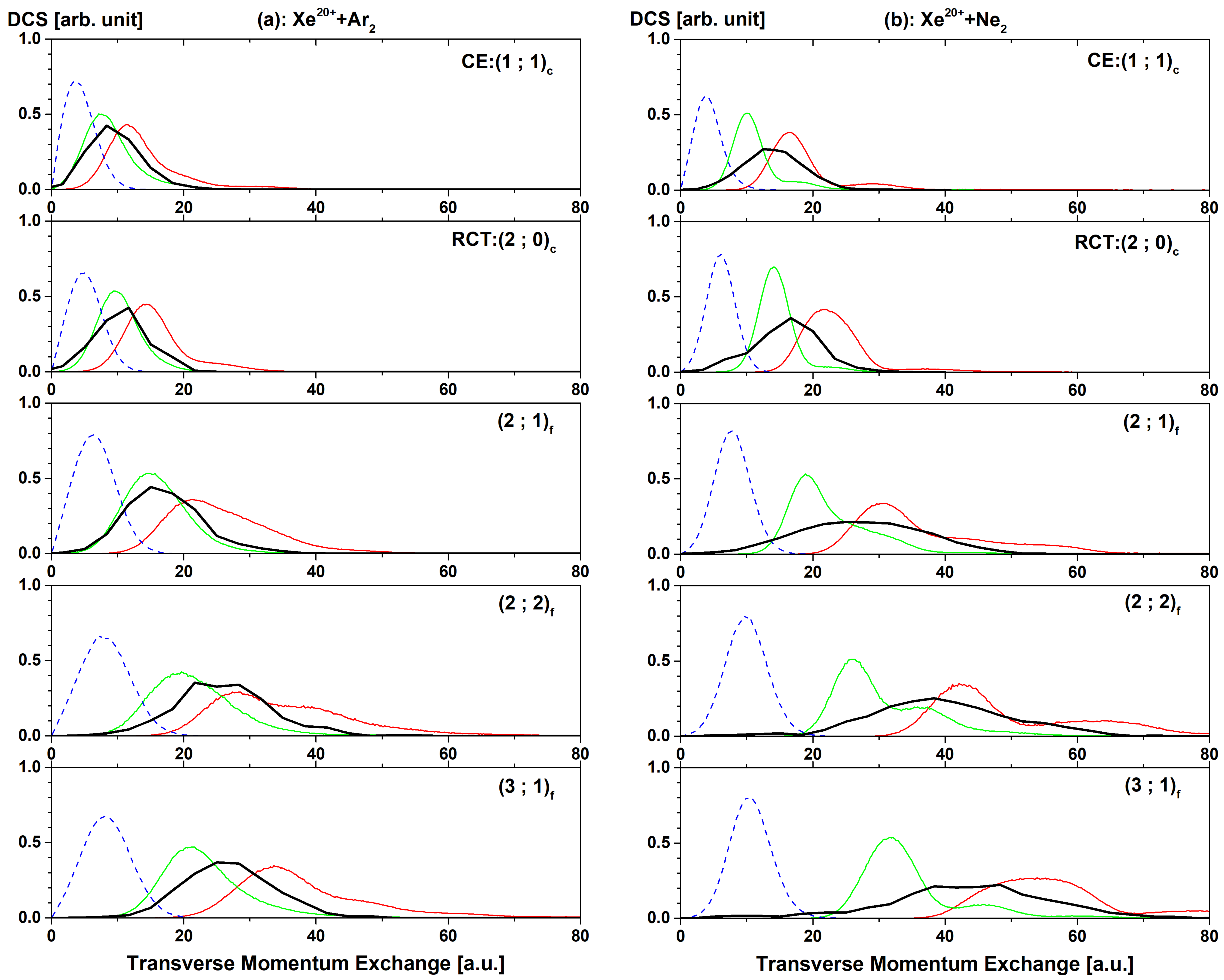}}
\caption{Differential cross section (DCS) in transverse momentum exchange obtained experimentally (thick black line) and using the MC-COBM for Xe$^{20+}$ projectiles colliding with Ar dimers (a) and Ne dimers (b). The MC-COBM distributions, previously normalized to the experimental data, are given for three different charge screening parameters: $S=1$ (blue dashed line), $S=1/2$ (green line (or gray)) and $S=0$ (red line (or dark gray)).}
\label{PCM_Xe}
\end{figure*}

For the experimental data, the transverse momentum exchange between the projectile and the center of mass of the dimer is inferred from the sum of the vector momenta of the two ionic fragments in the laboratory frame. For the MC-COBM ones, the momentum exchange induced by the collision is calculated as described in Sec.\ref{sec2-2}, using three different values of the charge screening parameter $S$ in Eq.\ref{qeff-in-dimer} and \ref{qeff-out-dimer}. The post-collision treatment of the data described in Sec.\ref{sec3-2} is then performed to account for the spatial extension of the collision region and the response function of the apparatus. As in \cite{Iskandar14}, the dimer orientation is selected with $\theta_B$ between 60$^{\circ}$ and 120$^{\circ}$ for both the experimental and the MC-COBM data, so that the dimer axis is quasi-perpendicular to the projectile beam axis. 

For all the processes and fragmentation channels discussed in the previous section, the transverse momentum exchange distributions obtained experimentally (thick black line) are compared to the the MC-COBM ones (color lines)  in Fig.\ref{PCM_Ar} for Ar$^{9+}$ projectiles and in Fig.\ref{PCM_Xe} for Xe$^{20+}$ projectiles. The MC-COBM distributions shown for the fragmentation channels (2;1)$_f$ and (3;1)$_f$ were obtained by summing the contributions of the (3;0)$_c$ and (2;1)$_c$ capture channels for (2;1)$_f$, and of the (4;0)$_c$ and (3;1)$_c$ capture channels for (3;1)$_f$, using the relative yields given by the gray columns of Fig.\ref{Ar-Ne-ratio} assuming a 50\% population of unbound states for one-site capture. Note that choosing the other scenario (light gray columns of Fig.\ref{Ar-Ne-ratio}, no unbound state population for one-site capture) lead to very similar distributions (not shown here). This can be explained by the weak dependence of transverse momentum exchange on the charge repartition within the ionized target, as can be seen when comparing the differential cross sections obtained for channels (1;1)$_c$ and (2;0)$_c$ in Fig.\ref{PCM_Ar} and Fig.\ref{PCM_Xe}.

The first MC-COBM momentum exchange calculation, with $S=1$ (blue dashed lines), assumes a full charge screening. Within this unrealistic model, the charge of shared electrons is allocated to the target effective charge. The second calculation, with $S=0$ (red or dark gray), corresponds to the picture of the collision given by Niehaus, with shared electrons acting as pure spectators without any screening contribution. The last one, with $S=0.5$ (green or gray), is an intermediate view assuming partial screening of the target by the shared electrons.

When assuming full screening of the target ($S=1$), the momentum exchange distribution is systematically strongly underestimated by the model. Considering shared electrons as pure spectators ($S=0$), as for the crossing radii and cross sections calculations, lead contrariwise to an overestimation of the transverse momentum exchange.
However, a target screening parameter $S=0.5$ reproduces remarkably well the experimental distributions. The mean values and the widths of the momentum exchange distributions, as well as their increase with the number of captured electrons are almost perfectly reproduced by the model for both systems Ar$^{9+}$+Ar$_2$ and Ar$^{9+}$+Ne$_2$. An excellent agreement is also obtained with $S=0.5$ for Xe$^{20+}$+Ar$_2$ collisions, where we only note a small underestimation of the mean momentum exchange for quadrupole electron capture. For Xe$^{20+}$+Ne$_2$ collisions, a screening parameter value of 0.5 seems too strong and lead to a small but systematic underestimation of the momentum exchange during the collision. Nevertheless, considering the extreme simplicity of the model, the overall agreement between its predictions and the experimental data remains very good for all systems when considering an empirical charge screening parameter $S$ close to 0.5 for the transverse momentum transfer calculation. It is in contradiction with the initial assumptions of the model of Niehaus used for the estimation of sharing and capture radii, where shared electrons are considered as pure spectators.

The effect of the target composition can also be discussed by comparing the results obtained for Ne dimers and Ar dimers (Fig.\ref{PCM_Xe}.a versus Fig.\ref{PCM_Xe}.b and Fig.\ref{PCM_Ar}.a versus Fig.\ref{PCM_Ar}.b). For both projectiles Xe$^{20+}$ and Ar$^{9+}$, the transverse momentum exchange produced with Ne dimers is larger than for Ar dimers. This is due to the different outer-shell number, $n=2$ for Ne atoms and $n=3$ for Ar atoms, of the electrons mostly involved in the charge exchange process. Ne dimers lead thus to smaller sharing and capture radii than Ar dimers. For a given capture channel, this results in smaller impact parameters and to larger Coulombic repulsion for Ne dimers, as observed experimentally and in the calculations.

\section{\label{sec7}SUMMARY}
An adaptation of the COBM model for low energy collisions between MCIs and rare gas dimer targets has been developed. It is based on the simple representation of the dimer as two atoms fixed in space and on the use of a MC simulation to integrate over the impact parameters and molecular orientations. An overall good agreement between the MC-COBM calculations and the experimental data has been obtained for the four collision systems investigated. Despite its simplicity, the model has provided for each channel of interest relative cross sections whose deviations with the experimental results remained below 10$\%$ of the total yield. The calculation of the transverse momentum exchange between the projectile and the target has also been implemented, providing a mean to study the effect of the electrons shared during the collision. A very good agreement has been obtained for a charge screening parameter $S=0.5$, indicating that shared electrons are not only spectators, as usually considered in the Niehaus model. The clear overall success of the present model shows that rare gas van der Waals dimers can be fairly well represented as two independent atoms. Further investigations with different charge states should help us to delimit the applicability of the model. 
In a close future, the same methodology could also be employed to investigate the collision dynamics involving more complex targets, such as larger homonuclear or mixed clusters.
\section*{ACKNOWLEDGMENTS}
We would like to pay a posthumous tribute to Dominique Hennecart, who played a major role in collision physics in Caen and whose work triggered the development of this MC-COBM model. We thank the CIMAP and GANIL staff for their technical support. This work was partly supported by TMU Research Program Grant.

\

\end{document}